\documentclass[aps,pre,onecolumn,showpacs,amsmath,amssymb,longbibliography,notitlepage,11pt,tightenlines]{revtex4-1} 
\usepackage{graphicx} 
\usepackage{dcolumn} 
\usepackage{bm} 

\begin{document}

\title{A topologically-protected interior for three-dimensional confluent cellular collectives}
\author{Tao Zhang$^1$}
\author{J. M. Schwarz$^{2,3}$}
\affiliation{$^1$Department of Polymer Science and Engineering, Shanghai Jiao
  Tong University, Shanghai 200240, PR China\\
 $^2$Physics Department, Syracuse University, Syracuse, NY 13244,
  USA\\
$^3$ Indian Creek Farm, Ithaca, NY 14850}



\begin{abstract}
Organoids are {\it in vitro} cellular collectives from which
brain-like, or gut-like, or kidney-like structures emerge. To make quantitative predictions regarding the
morphology and rheology of a cellular collective in its initial stages of development, we construct and
study a three-dimensional vertex model.  In such a model, the cells
are represented as deformable polyhedrons with cells sharing faces such
that there are no gaps between them, otherwise known as confluent. In a bulk model with periodic
boundary conditions, we find a  
rigidity transition as a function of the target cell shape index $s_0$ with a critical value $s_0^*=5.39\pm0.01$. For
a confluent cellular collective with a finite boundary, and in the presence of lateral extensile and in-plane, radial extensile deformations, we find a significant boundary-bulk effect that is one-cell layer thick. More specifically, for lateral extensile deformations, the cells in the bulk are much
less aligned with the direction of the lateral deformation than the cells at the boundary. For in-plane, radial deformations, the cells in the bulk exhibit much less reorientation perpendicular to the radial direction than the cells at the boundary. In other words, for both deformations, the bulk, interior cells are topologically-protected from the deformations, at least over time scales much slower than the timescale for cellular rearrangements and up to reasonable amounts of strain.  Our results provide an underlying mechanism for some observed cell shape patterning in organoids. Finally, we discuss the use of a cellular-based approach to designing organoids with new types of morphologies to study the intricate relationship between structure and function at the multi-cellular scale for example.
\end{abstract}

\maketitle

\noindent\textbf{\large Introduction}\\
\indent Organoids provide us with an {\it in vitro} window into
organogenesis~\cite{Lancaster2014,Clevers2016}. An organoid starts off as a cellular collective (a
clump of cells) consisting typically of cells, such as pluripotent stem cells.  The
cells are given some induction medium to help them differentiate into
either neurons or kidney cells or heart muscle cells.  The cellular
collective is then transferred to a petri dish containing Matrigel,
which contains collagen and, at times, growth factors and morphogens, followed by additional protocol, such as agitation~\cite{Lancaster2014b,Hofer2021}.  Even at the initial
stages, there is some patterning of the collective in terms of cell
shape~\cite{Karzbrun2018}. Over a longer time scale, as the cells divide, the organoid
continues to develop additional structure and a brain organoid~\cite{Lancaster2013},
for example, ultimately emerges. Intriguingly, brain organoids, exhibit the {\it in vivo} phenomenon of neuronal diversity~\cite{Velasco2019}. Additionally, at even later stages of development, neurons in the brain organoid
begin to fire and fire synchronously, just as in the developing
brain~\cite{Trujillo2019}. Intestinal organoids, on the other hand, self-assemble into crypt-villus units~\cite{Sato2009}.  In addition to providing a window into
organogenesis, organoids also provide an intermediary platform between cells and organs to study disease~\cite{Lancaster2014,Clevers2016}.  For instance, scientists are currently
exploring how SARS-CoV-2 infects the brain via brain organoid
studies to provide guidance for drug treatment of the COVID-19~\cite{Ramani2020,Zhang2020}.

We now hone in on the early stages of organoid development.  In particular, for quasi-two-dimensional
geometries, which are used in the formation of brain  organoids~\cite{Karzbrun2018}
and intestinal organoids~\cite{Gjorevski2022}, for instance, there appears to be differences in cellular structure between the cells in contact with the Matrigel at the edge of the organoid and the cells in 
the bulk (or interior).  In particular, the cells on the boundary appear to be
more elongated with the direction perpendicular to the edge of
the Matrigel.  The cells in the bulk, on the other hand, appear to be more globular with no apparent, particular orientation~\cite{Karzbrun2018}. For three-dimensional geometries, a central hole, or lumen, in place of a bulk of globular cells can emerge~\cite{Lancaster2013}.

Given the apparent ubiquity of the phenomenon, we ask: How does such a topologically-protected interior in these cellular
collectives arise? Theoretical work has been done to demonstrate how an individual versus multiple cortex-core structures materialize from a hydrodynamic description of active forces in cellular collectives to help regulate the overall brain organoid  architecture~\cite{Borzou2022}. Here, we implement a cellular-based computational approach to probe the above question, namely, a three-dimensional vertex model~\cite{Nagai1990,Fuchizaki1995,Honda2004,Okuda2015,Krajnc2018}.  Such models are ones
with cells represented as deformable polyhedrons and there are no gaps
between them. While two-dimensional vertex models have been studied widely~\cite{Honda1982,Farhadifar2007a,Staple2010a,Fletcher2014,Bi2015,Bi2016b,Yan2019,Sahu2020a,Kim2021,Luca2022},
three-dimensional vertex models have perhaps not been studied as extensively. Presumably,
this is, in part, because there is no publicly available code, while
two-dimensional codes, such as CHASTE~\cite{Fletcher2013,Mirams2013}, CellGPU~\cite{Sussman2017} and the different, but related,
Active Vertex Model~\cite{Barton2017}, are available. 

As for prior three-dimensional vertex model work that
has been done, researchers have found that introducing polarized interfacial tension allows for cells to migrate individually, or even as a
cluster, through a tissue~\cite{Okuda2021}. In addition, studies of branching in tissue
to form a lung have also been explored in such models~\cite{Okuda2018}. And yet, prior studies of the
three-dimensional vertex model have focused on a particular energy
functional that is different from a three-dimensional
extension of the two-dimensional energy functional in which a  rigidity transition
was uncovered~\cite{Bi2015} and with support from experimental observations of asthmatic bronchial epithelial tissue~\cite{Park2015}. More recently, the same two-dimensional vertex model extended to include two cell types predicted a new micro de-mixing phenomenon which was supported by experiments~\cite{Sahu2020a}. We will use
the same version of the energy functional as the one exhibiting a
density-independent rigidity transition in two dimensions to ask first whether or not there exists a
rigidity transition in a bulk model.  Prior work has demonstrated
that there exists a rigidity transition in a three-dimensional Voronoi
model as a function of the three-dimensional shape index~\cite{Merkel2018}. We will then explore how deformations of a confluent cellular collective
affect the morphology and the rheology of the collective. Our efforts represent a substantive leap from earlier, elegant work studying the shapes of shells composed of the polyhedrons as we now can probe the interior~\cite{Rozman2020}.

\bigskip
\noindent\textbf{\large Model}\\
\indent Cells are biomechanical and biochemical constructs that are not in
equilibrium, i.e., they are driven by internal, or active forces. The biomechanics
of the cellular collective is given by the energy functional:
      \begin{equation}
      \label{eq:energy}
      E= K_V\sum_{j}(V_{j} - V_{0})^2+K_A\sum_{j}(A_{j}
      -A_{0})^2 +\gamma\sum_{\alpha}\delta_{\alpha,B} A_{\alpha}, 
      \end{equation}
      where $A_{j}$ denotes the $j$th cell total area, the $j$th cell volume is denoted by
      $V_{j}$, and $\alpha$ labels the faces of the cells, with $\delta_{\alpha,B}=0$ if a face is not at the boundary B of the collective and 1 otherwise. 
      Given the quadratic penalty from deviating from a
      cell's preferred volume $V_0$ and area $A_0$, $K_V$ and $K_A$ are volume and
      area stiffnesses, respectively. Physically, the volume term represents the bulk elasticity
      of the cell with $V_0$ denoting a target volume. 
      The area term for cell $j$ can be rewritten as $K_A A_j^2 + \Gamma A_j + const$, 
      where $\Gamma =-2 K_A A_0$. Here the first term $K_A A_j^2$ represents the contractility of
      the acto-myosin cortex, and the second term $\Gamma A_j$ represents an interfacial tension $\Gamma$ set by a competition between the cell-cell adhesion at negative $\Gamma$ (larger $A_0$) and the cortical contractility at positive $\Gamma$ (smaller $A_0$). 
      Indeed, cell-cell adhesion and contractility are coupled~\cite{Hoffman2015}.  For instance, knocking out E-cadherin in keratinocytes, effectively changes the contractility~\cite{Sahu2020a}. We can tune for cell-cell adhesion through the target area $A_0$. The target area $A_0$ has a physical meaning of controlling whether the cell-cell adhesion or the cortical tension dominates. This translation from the mathematics to the biology is a generalization from prior translations for two-dimensional models in which the effective target perimeter $P_0$ (assuming $A_0=1$) translates to a competition between cell-cell adhesion 
      at larger $P_0$ and contractility at smaller $P_0$. Note that the target area $A_0$ is also related to the isotropy of cortical contractility. The larger the $A_{0}$, the less
      isotropically contractile the cell is, and vice versa. The less isotropically contractile a cell is, the more likely it can develop contractility in a particular direction, or anisotropic contractility as modulated by stress fibers, for example~\cite{Warmt2021}. As for the linear area term, the cells at the boundary of the cellular collective, there is an additional surface tension term for faces interacting with the ``vacuum''. One can nondimensionalize any length $l$ in the simulation with $l=V_0^{1/3}$. An important parameter in these models is the dimensionless shape index $s_{0,\beta} =A_{0,\beta}/(V_{0,\beta})^{2/3} $. A regular tetrahedron has a dimensionless shape index of $s_0\approx 7.2$, for example. 

\begin{centering}
\begin{figure}[tb]
  \includegraphics[width=20cm]{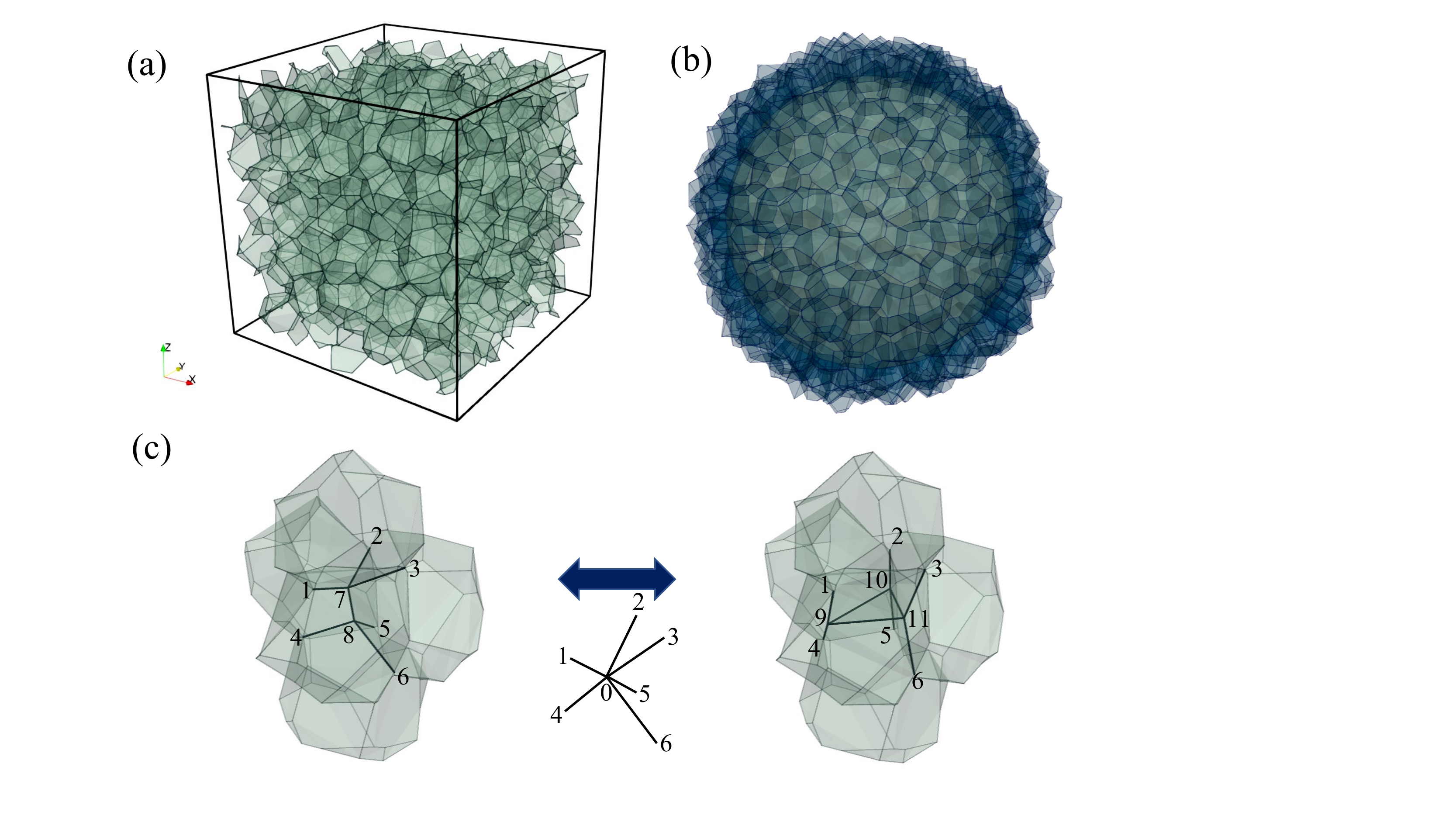}
  \caption{
  \textbf{3D vertex model} 
  \textbf{(a)} Snapshot of the bulk model with $s_0=5.60$.  
  \textbf{(b)} Snapshot of the cellular collective (light green) with $s_0=5.60$ and empty cells (dark green). 
  \textbf{(c)} Snapshot series of a reconnection event between cells in which the edge composed of the vertices labeled 7 and 8 becomes the triangle composed of vertices labeled 9, 10, and 11, and vice versa to result in a change of neighbors while remaining confluent. The numbers indicate serial vertices, and solid lines indicate edges. The point labeled 0 indicates the center of the reconnection event. 
  }
  \label{model}
\end{figure}
\end{centering}

Now that we have addressed the biomechanical aspect of cells. We must
also account for their dynamics. Cells can move past each other even
while the tissue remains confluent. In two dimensions such movements are
known as T1 events. Understanding such events are key to understanding
the rigidity transition in two dimensions~\cite{Bi2015}.   In three dimensions, such
movements are known as reconnection events. As for how cells exchange neighbors via a reconnection event, following Okuda and company \cite{Okuda2013}, we focus on edges going to triangles and vice versa.  See Figure 1(c). To determine if a reconnection event occurs, 
we look for edges with lengths less than $l_{th}$ and triangles with all three edge lengths less than $l_{th}$.  If there are indeed such edges or triangles, we choose one of the edges randomly and perform an edge-to-triangle reconnection event in which the edge vanishes and is
replaced by a triangle whose normal vector is parallel to the initial edge, or choose one of the triangles randomly and perform a triangle-to-edge reconnection event, should the following conditions introduced in the work of Okuda and company \cite{Okuda2013} be met to ensure whether or not a reconnection event is physically plausible. The {\it first condition} is that the change in energy before and after the reconnection event should be in the order of $l_{th}$. The {\it second condition} relates to resolving the topological irreversibility and satisfies the following sub-conditions: (1) two edges do not share two vertices simultaneously, (2) two polygonal faces do not share two or more edges simultaneously, and (3) two polyhedral cells do not share two or more polygonal faces simultaneously. The third sub-condition is implemented given the computational efficiency and was not discussed explicitly in the work of Okuda and company. 

After every ten time steps in the MD simulation are completed, all the edges and triangles that can undergo a reconnection event, given the above conditions, are tagged.  First, reconnection events from edge to triangle are performed sequentially. Second, reconnection
events from triangle to edge are performed sequentially. We allow for the possibility that some edges and triangles initially tagged can become untagged as the reconnection events occur.  Edges and triangles are only subtracted from the tagged list and not added to it. 

In addition to reconnection events, there is an underlying Brownian dynamics
for each vertex. Specifically, the equation of motion for the position ${\bf r}_I$ of a single vertex $I$ 
is 
\begin{equation}
\dot{\mathbf{r}}_{I}=\mu \mathbf{F}_{I} +\mu \mathbf{F}_{I}^B,\label{eq:dyn} 
\end{equation}
with $\mathbf{F}_{I}$ and $\mathbf{F}_{I}^B$ denoting the conservative force and
the random thermal force on the $Ith$ vertex respectively. 
The force $\mathbf{F}_I$ is determined from both the area and volume energetic constraints and, hence, includes cell-cell interactions. In addition, each $Ith$ vertex performs a random walk with an effective diffusion coefficient of $\mu k_B T$, where $T$ is an effective temperature. Unless otherwise specified, the mobility $\mu=1$.  Finally, the Euler-Murayama integration method is used to update the position of each vertex. See Table I for the listing of the parameters used in the simulations and their corresponding values. 

As for bulk case simulations, an initial state is created using a three-dimensional Voronoi
tessellation \cite{Rycroft2009} given randomized cell centers and assuming periodic
boundary conditions. The vertices, edges, and faces of each cell are then defined. 
To compute the force on each vertex due to the energetic contributions in Eq. 1, 
each polygonal face that has four or more edges is broken up into 
radially arranged triangles composed of each edge and the center point of the polygonal face \cite{Okuda2013}. 
The pressure of each cell due to the volume term and the tension of each triangular face due to the area term
in Eq. 1 are computed. 
The force on each triangular face is determined by multiplying the pressure with the surface area and 
multiplying the tension with the edge length. 
This force is then redistributed to the vertices making up each face. 
This method was implemented, as opposed to computing the partial
derivatives $\mathbf{F}_{I}=\nabla_I E$ 
of the energy $E$ with respect to the positions of the $Ith$ vertex  
given the computational efficiency of this latter method and
have checked that the two calculations are equivalent.
Once the forces are known, the positions of the vertices are updated using Eq. 2 that
includes a random thermal force $\mathbf{F}_{I}^B$. The simulation timestep is referred to as $d\tau$.
The system is equilibrated for a time period $\Delta t=10000$ before collecting data for the overlap function plots. 

To explore the properties of a cellular collective, one must consider
boundaries. For instance, as cellular collectives become embedded in Matrigel to ultimately
become organoids, they interact with the deformable Matrigel as they
develop~\cite{Lancaster2013}.  Here, we do not explicitly consider Matrigel, though prior work in two dimensions of a cellular collective embedded in a spring network has been done~\cite{Parker2020}.  Instead, we construct
a confluent cellular collective, or a clump of confluent cells, by
making a spherical cut-out of the bulk periodic
system that contains cells with empty cells beyond the boundary
between cells and empty space (see Figure 1(a) and (b))
as has been done in two dimensions~\cite{Parker2020}. For those cells at the boundary of
the clump, the interfacial vertices contain an additional
interfacial surface tension $\gamma$. Moreover, we allow reconnection events with more than one empty, or phantom, cells. 

Finally, we explore the rheological properties of the three-dimensional vertex
model in bulk as well as the confluent cellular collective.  For the
latter, we consider deformations on the cellular collective on the
boundary by having the boundary vertices undergo constant speed $v$
outward. In other words, a constant extensile
strain/deformation rate is imposed. In terms of direction, we explore two kinds of extensile
strain, one that is uni-axial and one that is in-plane, radial. One can
readily interpret the deformation as an external one, or one that is applied
to the clump.  Alternatively, one can interpret the deformation as an
internal one due to, for example, leader cells whose polymerization of
the cytoskeleton is in an outward direction much like a flexocytes
model with internal structure~\cite{Abaurrea2019}. 
Finally, these boundary vertices do not exhibit thermal-like fluctuations in their motion.

\begin{table}
\begin{center}
\begin{tabular}{ l c c }
 Diffusion constant & $D$ & $1$ \\ 
 Thermal energy & $k_BT$ & $10^{-4}$ \\   
 Simulation timestep & $d\tau$ & $0.005$ \\   
 Cell area stiffness & $K_A$& $1$ \\   
 Cell volume stiffness & $K_V$ & $10$ \\   
 Cell target volume & $V_0$ & $1$ \\ 
 Cell target surface area & $s_0$ & 5.0-5.8 \\  
 Boundary cell surface tension stiffness & $\gamma$ & 1 \\
 Boundary cell extension speed & $v$ & $10^{-4}$ \\  
 Reconnection event threshold edge length & $l_{th}$ & $0.02$ \\
 Number of bulk cells & $N_B$ & $512$ \\ 
 Number of cellular collective cells & $N$ & $~400$ \\
 Damping & $\xi$ & $1$\\  
 Number of realizations & $N_R$ & $20$\\
 Maximum strain of lateral extension & ${\epsilon}_l$ & $108\%$\\
 Maximum strain of radial extension & ${\epsilon}_r$ & $52\%$\\
 Long axis sample length after lateral extension & $L_l$ & $20$\\
 Inplane sample diameter after radial extension & $L_r$ & $14$\\
\end{tabular}
\caption{Table of the parameters used in the simulations.}
\end{center}
\end{table}

\bigskip
\noindent\textbf{\large Results}\\

\begin{centering}
\begin{figure*}[t]
  \includegraphics[width=15.5cm]{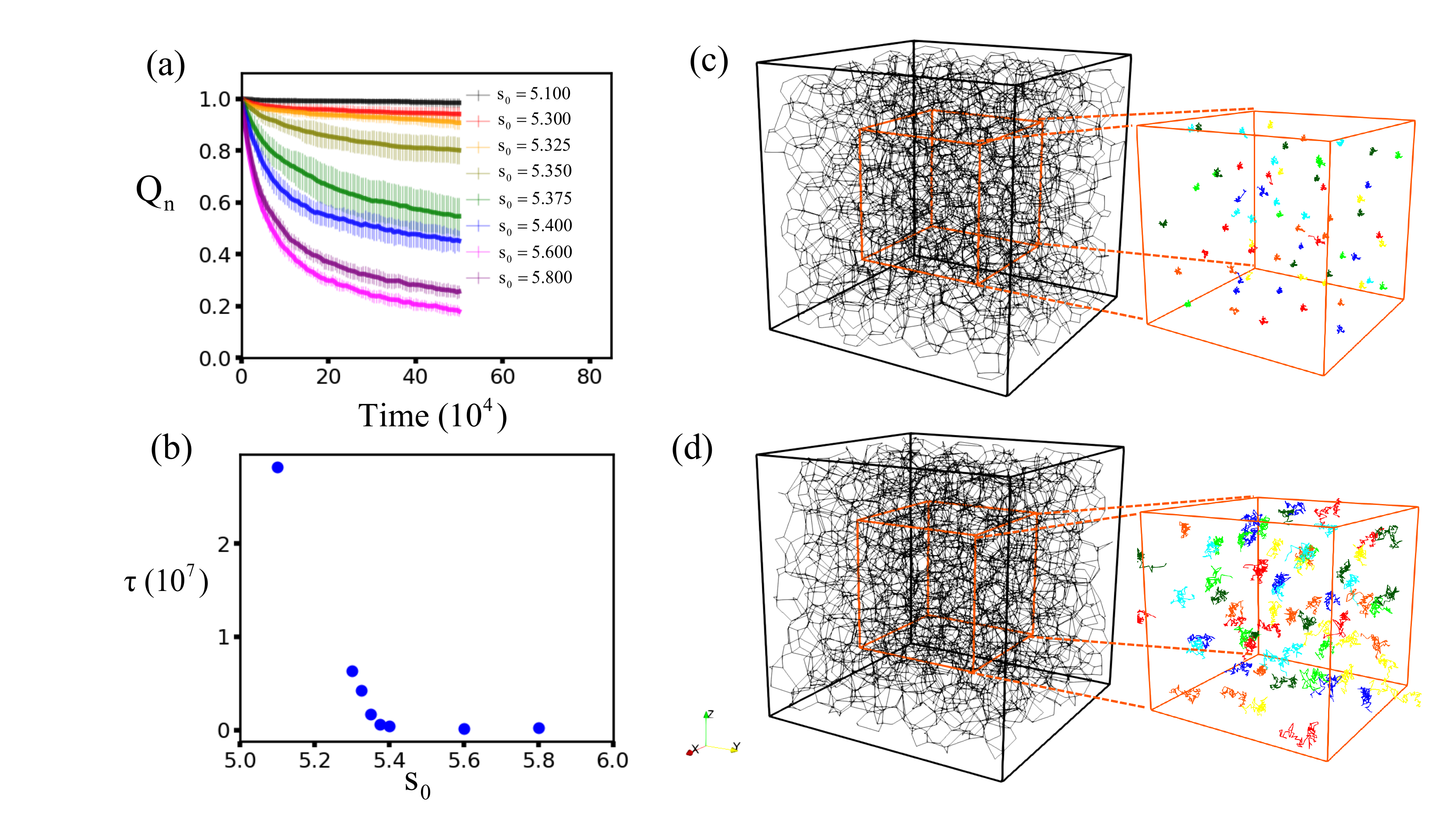}
  \caption{
  \textbf{Existence of a rigidity transition in the bulk model.} 
  \textbf{(a)} Overlap function $Q$ for different values of $s_0$. 
  \textbf{(b)} Floppy-rigid boundary: decay time $\tau$ as a function of $s_0$.
  \textbf{(c,d)} Trajectories of vertices initially located within the center cubic box with size 4.0 over a time period $\Delta t=400000$
  for $s_0=5.3,5.6$, respectively.
  }
  \label{phasediagram}
\end{figure*}
\end{centering}

\noindent\textbf{A rigidity transition in bulk.}
Prior studies have found a density-independent phase transition in
two-dimensional vertex models as well as a three-dimensional Voronoi
model with the same form of the energy functional \cite{Bi2015, Merkel2018}.  Is there
then a similar transition in the three-dimensional vertex model
presented here? To determine whether or not there is a transition, one can look at energy barriers to neighbor exchanges. In
this approach, one can determine a necessary 
condition for cellular rearrangement in two dimensions. For cells with area set to unity, for
example, having a target shape index that is regular pentagon involves no 
energy cost for 4-point vertex to emerge---the geometry at which
the cells perform a neighbor swap. Alternatively, one can measure the shear modulus of the system~\cite{Bi2015,Merkel2018}.  Since
we are implementing a dynamical approach, with each vertex undergoing
Brownian motion in a many-body, cell-cell interaction potential, one
can also determine whether or not the system is a fluid or solid by
looking at the mean-squared displacement of cells. However, it is
sometimes difficult to pinpoint the transition point given that the crossover time to a caging can be rather long~\cite{Bi2016b}. 

Instead, we use the concept of a neighbors-overlap function,
combined with trajectories, to determine whether or not the cells
remain localized or not~\cite{Giavazzi2018}. In particular, $Q_n$ is defined as
\begin{equation}
Q_n(t)=\frac{1}{N}\sum_{j=1}^{N} w_j,
\end{equation}
where $N$ is the number of cells, $w_j=0$ if cell $j$ has lost two or more neighbors and $w_j=1$ otherwise. Should the system be solid-like and so the cells do not
change neighbors, then $Q_n(t)=1$. Should the system become more fluid-like and
so cells do change neighbors, then $Q_n(t)<1$.  The smaller
$Q_n(t)$ becomes, the more cells change neighbors.  We must also
couple this measurement with observations of the trajectories of cells
since the changing of neighbors could result in some localized
movement, i.e., caging, as opposed to system-spanning trajectories, which are
indicative of something more akin to a fluid, as opposed to a glassy
state. Note that we have not yet incorporated a mechanism to prevent
back-and-forth reconnection events. Recent work implements such a
mechanism in two dimensions~\cite{Bi2021,Ganca2021} . 

We measure $Q_n(t)$ for the bulk system for different target shape
parameters, averaging over 20 realizations. See Figure 2(a).  We observe that for $s_0=5.1$, $Q_n(t)=1$ and remains at unity throughout the duration of simulation on
average. However, for $s_0=5.8$, it is clear that $Q_n(t)$ is decaying
to a value less than unity.  The overlap function is fit to an
exponential decay with $\tau_n$ defined as the decay
time. In Figure 2(b), we plot the decay
time $\tau_n$ as a function of $s_0$.  We find that $\tau_n$ approximates
zero around 
$s_0^*=5.39\pm0.01$, which is an estimate for the rigidity transition
point since the typical time scale for cellular rearrangements becomes
zero. Additionally, some cell trajectories are plotted in Figures 2(c) and (d). Some trajectories span larger than one cell length, 
for the larger $s_0$, to indicate that the cells are not caged, at least on that scale, and so not
simply switching back and forth between neighbors. We, therefore, find that both solid-like
and fluid-like behaviors as a function of $s_0$, suggesting a density-independent rigidity transition in this 
three-dimensional vertex model. We also study the cell shape index distribution to find that for $s_0=5.40$ and above the average of the distribution tracks the target $s_0$ indicating that cells can achieve their target shape while in the fluid-like state, while in the solid-like state, the average of the distribution cannot track the target $s_0$ and the distribution is more broaden than the fluid-like state (See Supplemental Material Figure S1~\cite{SM}). We note that if we extend the definition of $w_j$ to be $w_j=0$ if cell $j$ has lost $N_l$ or more neighbors and $w_j=1$ otherwise, where $N_l=3,4$, the rigidity transition point remains unchanged (See Supplemental Material Figure S2~\cite{SM}). 

\begin{centering}
\begin{figure*}[t]
  \includegraphics[width=16.4cm]{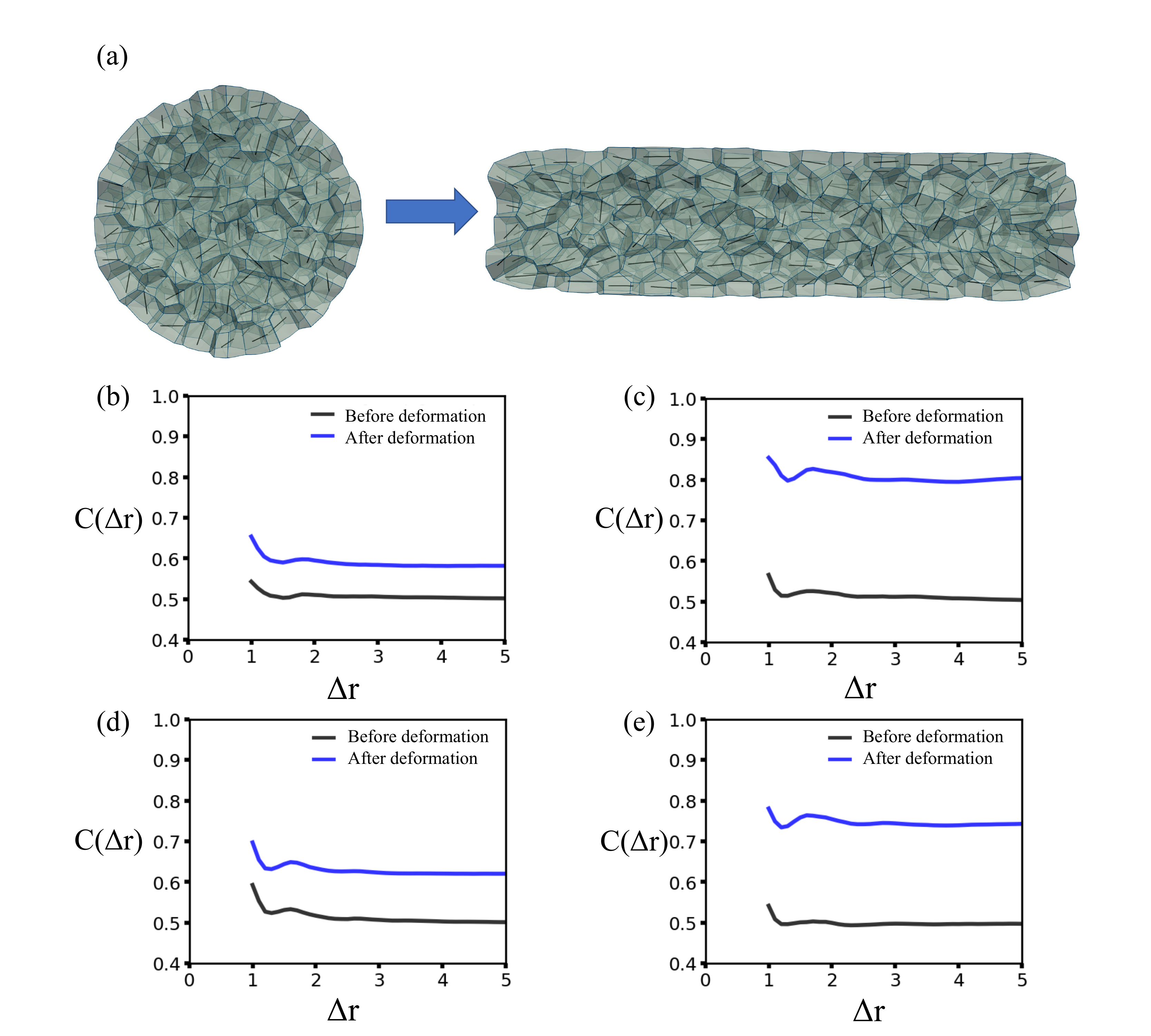}
  \caption{
  \textbf{Boundary-bulk morphology with lateral extensile deformation.}
  \textbf{(a)} Cross-sectional snapshots of the cellular collective before and after the lateral  extensile deformation, 
  where the black colored rods indicate the orientation of cells. 
  \textbf{(b,c,d,e)} The correlation function $C(\Delta r)$ as a function of the distance $\Delta r$ between the centers of two cells 
  before and after the lateral extensile deformation 
  for (b) bulk cells with $s_0=5.6$; (c) boundary cells with $s_0=5.6$; (d) bulk cells with $s_0=5.0$; (e) boundary cells with $s_0=5.0$.
  }
  \label{lateral_extension}
\end{figure*}
\end{centering}

\medskip
\noindent\textbf{Boundary-bulk morphology with lateral extensile strain.} 
While understanding additional bulk behavior in three dimensions is important, we now focus on a
confluent cellular collective, or a clump of cells. As discussed above, we take the
initial Voronoi construction (based on randomized points in
three dimensions) and then carve out a cluster of cells such that any
polyhedron that is not part of the cellular collective is simply empty
space. The faces of cells in contact with empty have an extra
energetic interfacial surface area contribution.  

To explore the structure and rheology of the cellular collective as it undergoes
a deformation, we assign the vertices in contact with ``empty space'' a
deterministic velocity $v$ outward along a particular, uni-axial
direction. As these vertices move outward, the structure of the
cellular collective adjusts so that the collective
remains confluent. Moreover, we explore whether or not a
collective that begins as a solid with $s_0=5.0$ and then becomes a
fluid as a result of the deformation. In other words, how do
deformations affect the rheology of the cellular collective?

As for the structure of the cellular collective following
the uni-axial, extensile deformation, we ask whether or not cells align along the
uni-axial direction of extension. Prior studies of epithelial colonies
that are uni-axially pulled are able to identify topological defects
within a nematic background~\cite{Comelles2021}. Such defects are recapitulated by a
two-dimensional vertex model with additional polarization terms added
to energy functional to account for polarization, and nematic ordering is present even before the pulling~\cite{Comelles2021}. While
we do not observe nematic ordering before the uni-axial, extensile  deformation, we ask
whether or not there exists emergent nematic ordering along the
uni-axial direction.  To test for this, we fit each polyhedron to a
minimal volume ellipsoid and then determine the orientation of the long
axis, dub it a ``rod'', and compute the average of the absolute value of the cosine of the
angle between any two rods for any pair of rods $(i,j)$ some distance between the two centers $r_{ij}\le\Delta r$, or 
\begin{equation}
C(\Delta r)=\frac{1}{M}\sum_{r_{ij}\le\Delta r}
\frac{|\vec{d}({\bf r})\cdot \vec{d}({\bf r}+r_{ij}
{\bf n})|}{|\vec{d}({\bf r})||\vec{d}({\bf r}+r_{ij}{\bf n})|}, 
\end{equation}
where $M$ is the
number of pairs, ${\bf n}$ is the three-dimensional unit vector, and $\vec{d}$ is the unit vector aligned with the rod to represent the rod orientation. The correlation function $C(\Delta r)$ indicates
the correlation in orientation between two rods with the centers a
distance $r\le\Delta r$ from each other.

Since the vertices on the edge of the cell collective are different from
the bulk cells in that there is an extra energetic contribution, we
separate out the boundary cells from the bulk cells and look for
correlation in alignment between cells in the bulk and between cells
on the boundary.  For $s_0=5.6$, we find that for bulk
cells, the spatial correlation in alignment is rather similar to 
the bulk system with periodic boundary conditions (See Supplemental Material Figure S3~\cite{SM}). 
In other words, we do not observe nematic ordering.  The boundary cells,
however, do exhibit ordering with a much higher correlation in
alignment as compared to the bulk cells. Intriguingly, the drop in alignment correlation from the
boundary to the bulk is
significant given that the boundary is only one-cell layer
thick. See Supplemental Material Figure S4~\cite{SM} for the plots of $C(\Delta r)$ for the outermost layers and the second outermost layer for additional support of this statement. In other words, there is a one-cell skin depth phenomenon in which the boundary cells
align with the deformation, but the bulk cells do not. While this may not be unexpected for a fluid-like system, this difference emerges in the solid-like system
with $s_0=5.0$ as well, however, the difference in spatial correlation
in alignment amongst cells in the bulk and amongst cells in the
boundary, is not as large. Presumably, this trend is because 
there are fewer reconnection events in the solid-like phase, so the
system overall is less plastic in the presence of the deformation and,
therefore, less responsive to the deformation. See Supplemental Material Figure S5~\cite{SM} for the plots of the distribution of the aspect ratio of boundary/bulk cells for additional support of this one-cell skin depth phenomenon. During the lateral extension, the boundary cells undergo much more aspect ratio change than the bulk cells in both fluid-like and solid-like systems, showing that the boundary cells can mechanically protect the interior. 

What is the mechanism driving this one-cell layer skin depth 
phenomenon? Earlier two-dimensional studies found a sharp, but deformable,
boundary between two groups of cells
distinguished solely by an interfacial line tension between
them~\cite{Sussman2018b}. More specifically, a discontinuity in the restoring force once a cell has ``invaded''
the territory of the other group of cells was determined~\cite{Sussman2018b}.  More recent work with
three-dimensional Voronoi models demonstrates that there is an
energetic barrier to destabilizing the interface between two cells
types within a layered geometry and that within a two cell layer
geometry, the centers of the cells align~\cite{Sahu2021}. Here, we observe that the boundary cells are rather responsive to the deformation along with smoothly connected boundary faces given the interfacial surface tension. However, their inner faces can take on a more jagged surface in the absence of an inner interfacial surface tension between the boundary cells and the second inner layer of cells.  With this more jagged surface, the interior cells remain insulated, or topologically-protected, from deformations at the surface. Note that this behavior is also distinct from earlier work in which a bulk two-dimensional vertex model undergoes pure shear, which results in alignment of the cells~\cite{Wang2020}.  Here, we allow the faces of the system to deform such the overall cellular collective shape is not constrained and includes curvature such that the cells take on a more varied zoology of shapes beyond the one-sided right prism~\cite{Sahu2021}. Moreover, the difference in organization between the boundary cells and the bulks cells persists for faster deformations, though the difference decreases particularly for the fluid-like case (See Supplemental Material Figure S6~\cite{SM}). 

As for the rheology in response to the deformation, we find the
distribution of the shape index of individual cells broadens for both
the fluid-like case and the solid-like case, particularly amongst the boundary cells (See Supplemental Material Figure S7~\cite{SM}). Interestingly, for the solid-like case, the cell shape index for the boundary cells becomes more broad than for the fluid-like case as the bulk cannot reorganize in response to the deformation, and so the boundary cells remodel even more so in response. In both cases, the enhancement of the cell shape index, particularly in the boundary cells, is due to a decrease in the volume.  Variations in the relative stiffnesses of $K_A$ and $K_V$ will modulate this response with a larger $K_V$ resulting in more changes in surface area than in volume of the boundary cells. 

In the solid-like system with $s_0=5.0$, the decrease in alignment correlation from the boundary to the bulk decreases as the magnitude of the additional surface tension of boundary cells $\gamma$ is decreased to $\gamma=0.5$, and increases as $\gamma$ is increased to  $\gamma=2.0$ (See Supplemental Material Figure S8~\cite{SM}). Due to fewer reconnection events in the solid-like phase, an increase in $\gamma$ leads to more volume shrinkage of boundary cells than the bulk cells, that is a larger volume difference between boundary and bulk cells. As a result of the larger volume difference along with a larger surface tension difference between the outer and inner faces of the boundary cells, the bulk-boundary difference in alignment correlation is enhanced. Similarly, a decrease in the cell volume stiffness $K_V$ leads to larger (boundary surface tension induced) volume shrinkage of boundary cells, so that the bulk-boundary difference in alignment correlation increases (See Supplemental Material Figure S9~\cite{SM}). However, in the fluid-like system with $s_0=5.6$, with more reconnection events, both boundary and bulk cells experience volume shrinkage together due to an increase in $\gamma$, so the bulk-boundary difference in alignment correlation will not increase. Moreover, while the average volume of bulk cells is further decreased, the target surface area of each bulk cell is still $A_0=5.6$. Hence the bulk cells intend to have an elongated shape to meet the target surface area, which results in a slight increase in alignment correlation. The decrease in alignment correlation from the boundary to the bulk in fluid-like system also shows a different trend than the solid-like system as $K_V$ is varied. An increase in $K_V$ leads to less volume change but more surface area change in bulk cells, so the bulk cells are able to reconfigure their shapes more freely during later extension, that is a lower alignment correlation and a larger decrease in alignment correlation from the boundary to the bulk.

\begin{figure*}[t]
  \includegraphics[width=14.5cm]{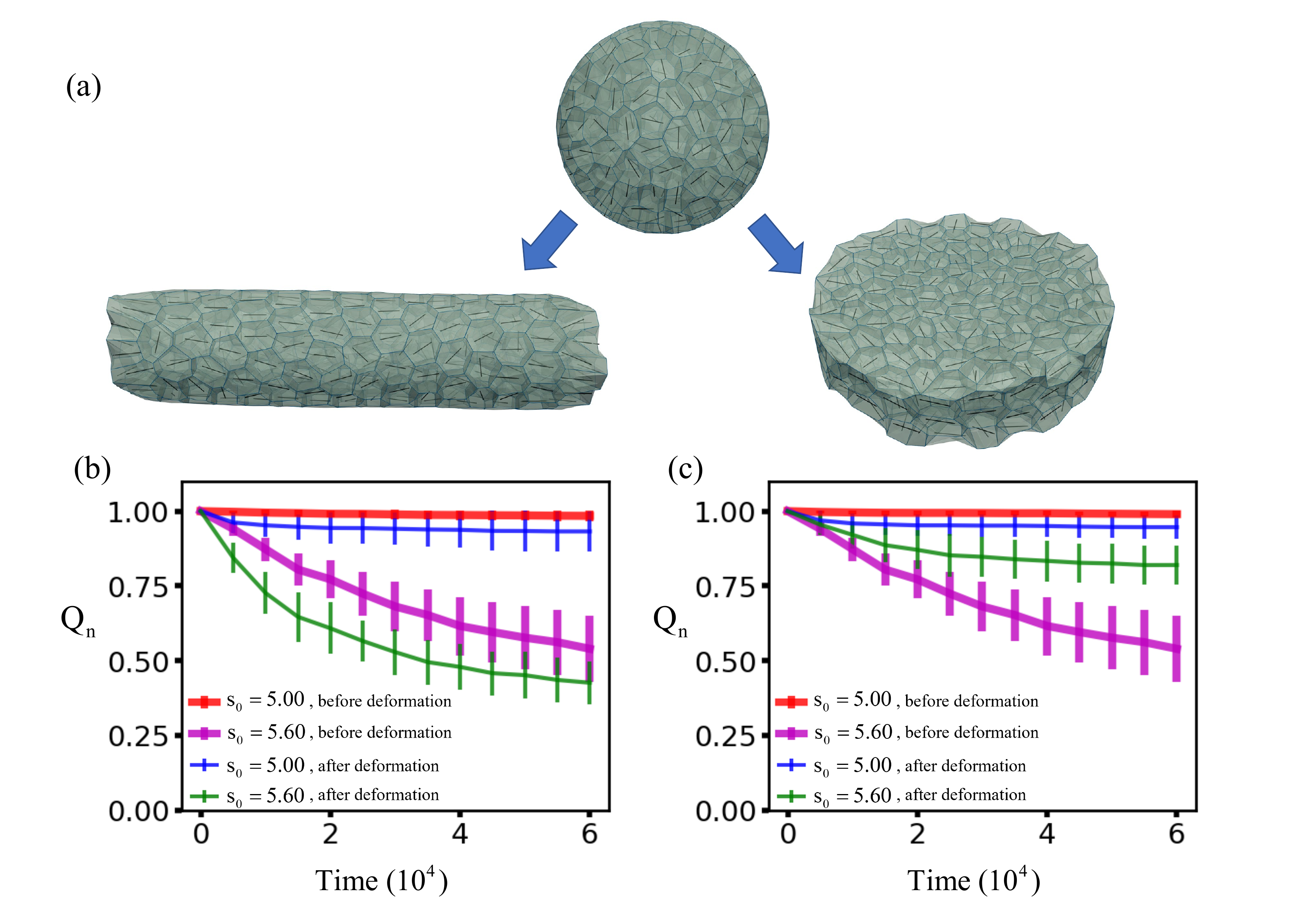}
  \caption{
  \textbf{Rheology response to deformation.} 
  \textbf{(a)} Snapshots of the cellular collective before and after the lateral and radial extension, 
  where the black colored rods indicate the orientation of cells. 
  \textbf{(b,c)} The overlap function $Q_n$ of the bulk cells as a function of simulation time
  with $s_0=5.0, 5.6$, 
  before and after (b) lateral extension and (c) radial extension.
 }
  \label{fluidity_extension}
 \end{figure*}

One might initially surmise that
with the increase in the individual cell shape index, the cell collective
becomes more fluid-like in response to the deformation in both
cases. However, with more of the changes occurring in the boundary cells, some of which were initially bulk cells at the start of the deformation, it is not clear. Measurements of the overlap function indicate that for
$s_0=5.6$, the cellular collective does not become more fluid-like with
similar decay of the overlap function $Q_n$ (as a function of time) before and after the deformation. See Figures 4(a) and (b). Moreover, 
for $s_0=5.0$, there also does not appear to be enhanced fluidization of the
system. This finding is consistent with the bulk of the cells remaining topologically-protected from the deformations at the boundary. However, it would be interesting to determine if the shear thinning observed in a bulk two-dimensional system~\cite{Duclut2021} occurs here.\\

\medskip
\noindent\textbf{Boundary-bulk morphology with in-plane, radial extensile strain.}  
Now that we have analyzed the confluent cellular collective's response to lateral, 
extensile strain, we explore its response to in-plane, radial  
extensile strain. Given the sharp boundary patterning in the prior 
case, we expect the bulk cells to neither align in some arbitrary direction.  We find that the nematic alignment correlation function for the bulk cells behaves similarly to the lateral extensile deformation. In other words, there is little alignment in the bulk. Given the
radial geometry, however, one may anticipate other types of
structural ordering, such as those observed in liquid crystals. For
instance, splay, or $S_i({\bf r})=
\frac{|\vec{\nabla}\cdot\vec{d}_i({\bf r})|^2}{|\vec{d}_i({\bf r})|^2}$ may be relevant. Again,
the denominator is present since $\vec{d}_i({\bf r})$ differs in length
from cell to cell. While there is very recent elegant work quantifying two-dimensional vertex models as $p$-actic liquid
crystals~\cite{Luca2022}, we will take a simpler
approach. We compute for each cell,  
\begin{equation}
\eta_i({\bf r})=|\cos(\theta_i({\bf r}))|=\frac{|\vec{n}_{\bf r}\cdot\vec{d}_i({\bf r})|}{|\vec{d}_i({\bf r})|}.  
\end{equation}
and plot the distribution of
angles $|\cos\theta_i({\bf r})|$. Here $\vec{n}_{\bf r}={\bf r}/|{\bf r}|$. 
Should $|\cos(\theta_i({\bf r})|\approx 0$
then the cells exhibit more curl than splay; should
$|\cos(\theta_i({\bf r})|\approx 1$, then the cells exhibit more splay than curl. Note that it is easier
to compute $|\theta_i({\bf r})|$ than the curl due to numerical inaccuracies. See Figure 5. We find that
before the deformation, splay is maximized at the
boundary for both $s_0=5.6$ and $s_0=5.0$. However, after the deformation, the boundary cells are oriented circumferentially to exhibit curl. On the other hand, the orientation of the bulk with respect to the radial direction does not change significantly.  So, again, with this alternate type of deformation, we observe a boundary-bulk patterning in which the interior cells remain topologically protected. Note that the patterning is different than those observed in quasi-two-dimensional brain organoids, where the boundary cells were oriented radially~\cite{Karzbrun2018}. Bulk cells are indeed moving to the boundary as the deformation occurs, but given the deformation rate, bulk cells are not moving to the boundary fast enough.  Should more bulk cells move to the boundary during the deformation, we hypothesize that the boundary cells may become oriented radially. In addition, the boundary-bulk patterning persists for faster deformation rates, while smaller interfacial surface tensions, it is not as robust. 

\begin{figure*}[t]
  \includegraphics[width=14.5cm]{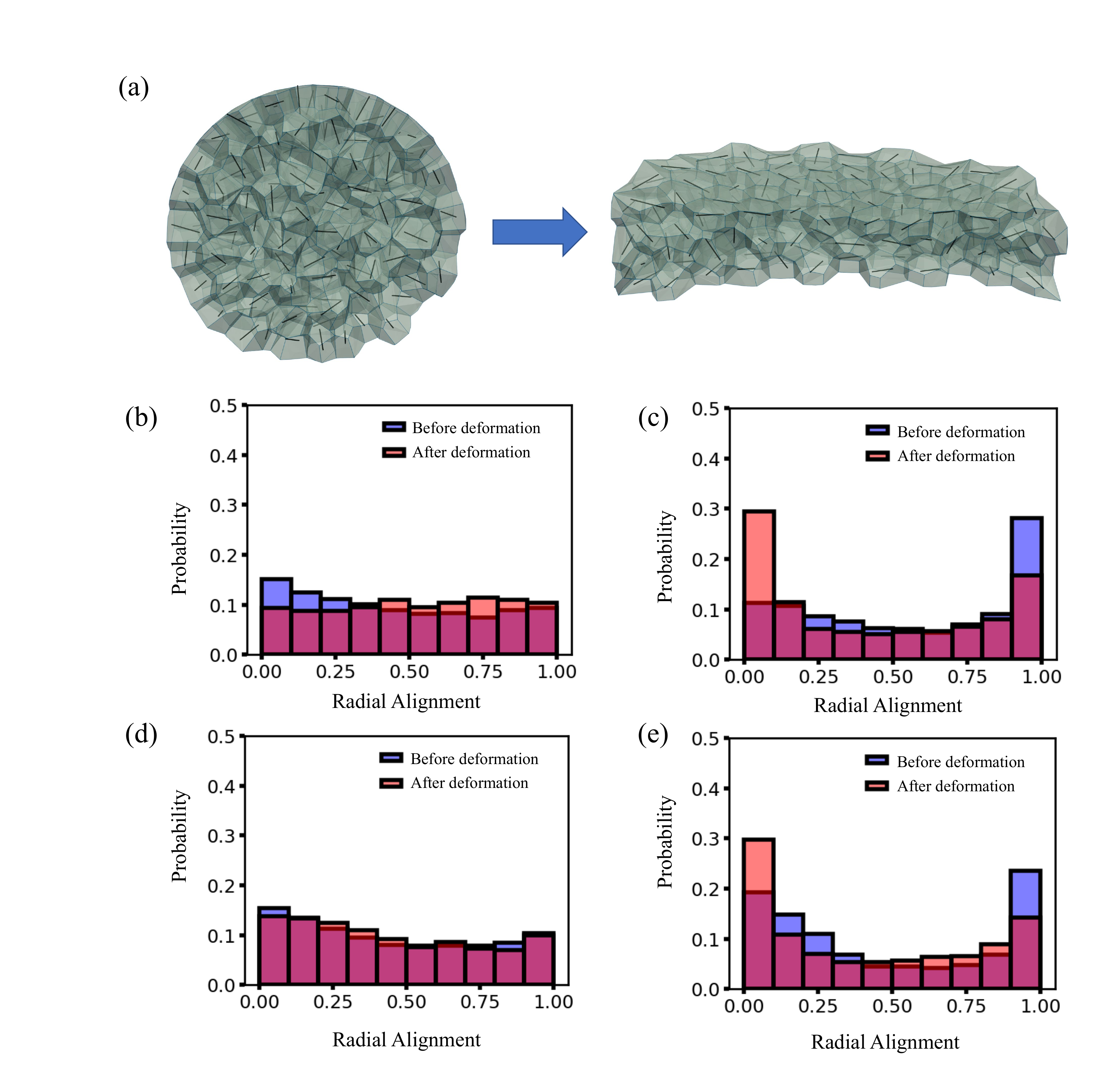}
  \caption{
  \textbf{Boundary-bulk morphology with in-plane, radial extension.} 
  \textbf{(a)} Cross-sectional snapshots of the cellular collective before and after the radial extension, 
  where the black colored rods indicate the orientation of cells. 
  \textbf{(b,c,d,e)} The distribution of $\eta_i({\bf r})$ before and after the in-plane, radial extension 
  for (b) bulk cells with $s_0=5.6$; (c) boundary cells with $s_0=5.6$; (d) bulk cells with $s_0=5.0$; (e) boundary cells with $s_0=5.0$.
 }
  \label{radial_extension}
 \end{figure*}

We also investigate how the confluent cellular collective rheology is in response to the
in-plane, radial extensile deformation. We observe similar trends as in the lateral extension case in which the solid-like system remains solid-like in terms of the
overlap function after the extension. However, for the fluid-like case, the system becomes more solid-like. Intriguingly, this result appears to be in-line with earlier work demonstrating compression-induced fluidization such that one might expect extension-induced solidification~\cite{Parker2020}. Presumably, the bulk cells in the cortex-core structure for the quasi-two-dimensional brain organoids are also somewhat solid-like~\cite{Karzbrun2018}. Changes in the cell shape index distribution before and after the deformation appear to be similar to the lateral extension, despite the change in rheology in the fluid-like case to become more solid-like (See Supplemental Material Figure S10~\cite{SM}). Again, much of the changes in shape index are focused at the boundary of the confluent cellular collection.

\bigskip
\noindent\textbf{\large Discussion}\\
\indent We construct and study a three-dimensional vertex model with a quadratic energy
functional in terms of a target surface area and a target 
volume using over-damped Brownian dynamics simulations. In doing so, we uncover a rigidity transition in the bulk
system with periodic boundary conditions. By measuring a discrete neighbor
overlap function, we determine the transition location to occur at a target shape index of  $s_0^*=5.39\pm0.01$, for the system
sizes and temperatures studied.  This transition location 
is slightly lower than the location of the rigidity transition
observed in a three-dimensional Voronoi version, where the degrees of freedom are assigned to the cell centers, using energy minimization~\cite{Merkel2018}. Density-independent fluidization transition occurs as isotropic contractility decreases so that anisotropic contractility via stress fibers may ultimately drive cell motion. Such an effect in single cells has been recently emphasized~\cite{Warmt2021} and is likely to occur at the multi-cellular level~\cite{Ilina2020}.  

We have also gone beyond a bulk system to examine confluent cellular
collectives and their response to deformations. We indeed observe
vestiges of the rigidity transition in the confluent cellular
collective case. For both lateral and in-plane radial extension, we observe
larger changes in the patterning before and after the deformation for the fluid
case. Specifically, the change in the nematic-like, or curl-like, 
alignment, of the boundary cells is greater in the fluid case. Specifically, for the lateral extension, the cells align with
the direction of the extension, while for in-plane radial extension, the
boundary cells align perpendicularly to the radial direction. Importantly, there is a significant difference
between the bulk cells and the boundary cells in terms of arrangement for both
types of deformations. In particular, the bulk cells resemble the cells in the bulk system with periodic boundary conditions, as if there were no boundary deformation at all. In other words, the bulk cells are topologically protected from deformations at the boundary.  

As for existing numerical evidence for a boundary-bulk effect, Finegan et al.~\cite{Finegan2019} details a wonderful experiment-theory collaboration addressing how Sidekick drives a particular type of reconnection event in Drosophila during a particular time in its development. While the experimental data seems to focus mostly on the bulk, an image of a two-dimensional vertex model simulation with a boundary shows the cells at the boundary do appear to more aligned with respect to each other as compared to alignment in the bulk. As for experimental evidence for a boundary-bulk effect, we refer to an image of the presomitic mesoderm of zebrafish embryo in Mongera et al.~\cite{Mongera2018}. While the boundary in the image is not between cells and “empty space” (fluid), but between effectively two different types of cells, ones that are thicker and ones that thinner, such that there is interfacial tension between the two types, one can observe a “top” row of very prism-like cells (in the thicker cells) and just below this “top” row exists cells that are not prism-like and not very ordered. In other words, we have some experimental evidence for the boundary-bulk effect, though such an effect needs to be carefully quantified experimentally. 

The stark difference between the bulk and boundary cells is a
phenomenon that cannot be readily captured in a continuum model. And while there are other ways to depict three-dimensional cellular collectives, such as cellular Potts model~\cite{Glazier2005} or a three-dimensional Voronoi model~\cite{Merkel2018,Grosser2021}, the  three-dimensional vertex model is our model of choice. 
In a recent theoretical study on cell extrusion in planar epithelial~\cite{Okuda2020}, 3D vertex model is shown to be capable of modeling scutoids-like packing of epithelia~\cite{Gomez2018} with an extra vertex appearing along the apico-basal axis. 
The topologically-protected interior is a consequence of the absence of any interfacial surface tension at the inner faces of the boundary cells. Moreover, the overall shape of the confluent cellular collective is deformable.  With these features, the cells can take on a more varied zoology of shapes beyond right prisms observed a two-layered, two cell-type Voronoi model with interfacial surface tension between the two cell types~\cite{Sahu2021}.  
In terms of timescales, at least over time scales larger than the time scale for cellular rearrangements, the boundary cells are indeed insulating the bulk cells from the boundary deformation as the confluent cellular collective deforms. Over shorter times, the shape of the cells mimic the boundary deformation, just as an elastic solid. One can therefore observe in a developmental system, for instance, different regimes of deformations at the cell scale in Drosophila epithelial morphogenesis~\cite{Finegan2019}.

The single-cell-layer thick boundary effects provide a mechanism
for patterning with a confluent cellular collective in which its interior remains topologically-protected from deformations at the boundary. For instance,
quasi-two-dimensional brain organoids have a cortex that is approximately one-cell layer
thick~\cite{Karzbrun2018}. And while in our system, the cells orient circumferentially at the boundary after in-plane, radial extension, modulation of strain rate and deformation protocol may effect the orientation of the cells at the boundary. We leave this for future work. 
If we can understand the mechanisms behind structure formation in
organoids, more generally, we can better control their formation so that it becomes more
``deterministic''.  This cell-based approach can then ultimately be used to recapitulate more complex organoid structures, such an octopus-like  brain with a core brain and its eight mini-brains~\cite{Hochner2012}. While evolution has selected for structures that optimize for specific functions given physical constraints, we can ultimately design brain organoid morphologies, for example, that will allow us to probe the intricate structure-function relationship at the multi-cellular scale. 

Even with this simple confluent cellular collective model that does not investigate additional types of interactions between cells such as polarized
tension~\cite{Comelles2021,Okuda2021}, we observe nontrivial phenomenon. It would be interesting to include cell growth, the opening of holes~\cite{Kim2021}, polarized interactions between cells~\cite{Comelles2021,Okuda2021}, and strain-dependent tension remodeling~\cite{Staddon2019,Cavanaugh2020}.  Moreover, with cell growth in a confined, but deformable, environment, such as Matrigel, compressive deformations also may arise.  But before building too complex a model, which is typically done to recapitulate experiments, we must first understand the most minimal models at hand, as they themselves can exhibit rich behaviors that the biology has discovered long before our own brains have.

\bigskip
\noindent\textbf{\large Data availability}\\ 
The data that support the findings of this study are available upon request. \\
\noindent\textbf{\large Acknowledgments}\\
We acknowledge discussion with Tara Finegan, Sarthak Gupta, Lisa
Manning, and Paula Santamsu. JMS acknowleges financial support from NSF-PoLS-2014192. We also acknowledge helpful feedback from an anonymous reviewer. \\
\noindent\textbf{\large Author contributions}\\
TZ and JMS conceived the study; TZ wrote the code and conducted the simulations; TZ and JMS interpreted the results and wrote the paper. \\
\noindent\textbf{\large Corresponding Authors}\\
Tao Zhang and J. M. Schwarz.\\
\textbf{\large Competing interests}\\
 The authors declare no competing financial or non-financial interests. \\

\bibliography{3Dvertex}

\newpage

\begin{widetext}
\begin{center}
\textbf{\large A topologically-protected interior for three-dimensional confluent cellular collectives\\ Supplemental Material}
\end{center}

\setcounter{equation}{0}
\setcounter{figure}{0}
\setcounter{table}{0}
\setcounter{page}{1}
\makeatletter
\renewcommand{\theequation}{S\arabic{equation}}
\renewcommand{\thefigure}{S\arabic{figure}}
\renewcommand{\bibnumfmt}[1]{[A#1]}
\appendix

\begin{figure*}[h]
  \includegraphics[width=14.5cm]{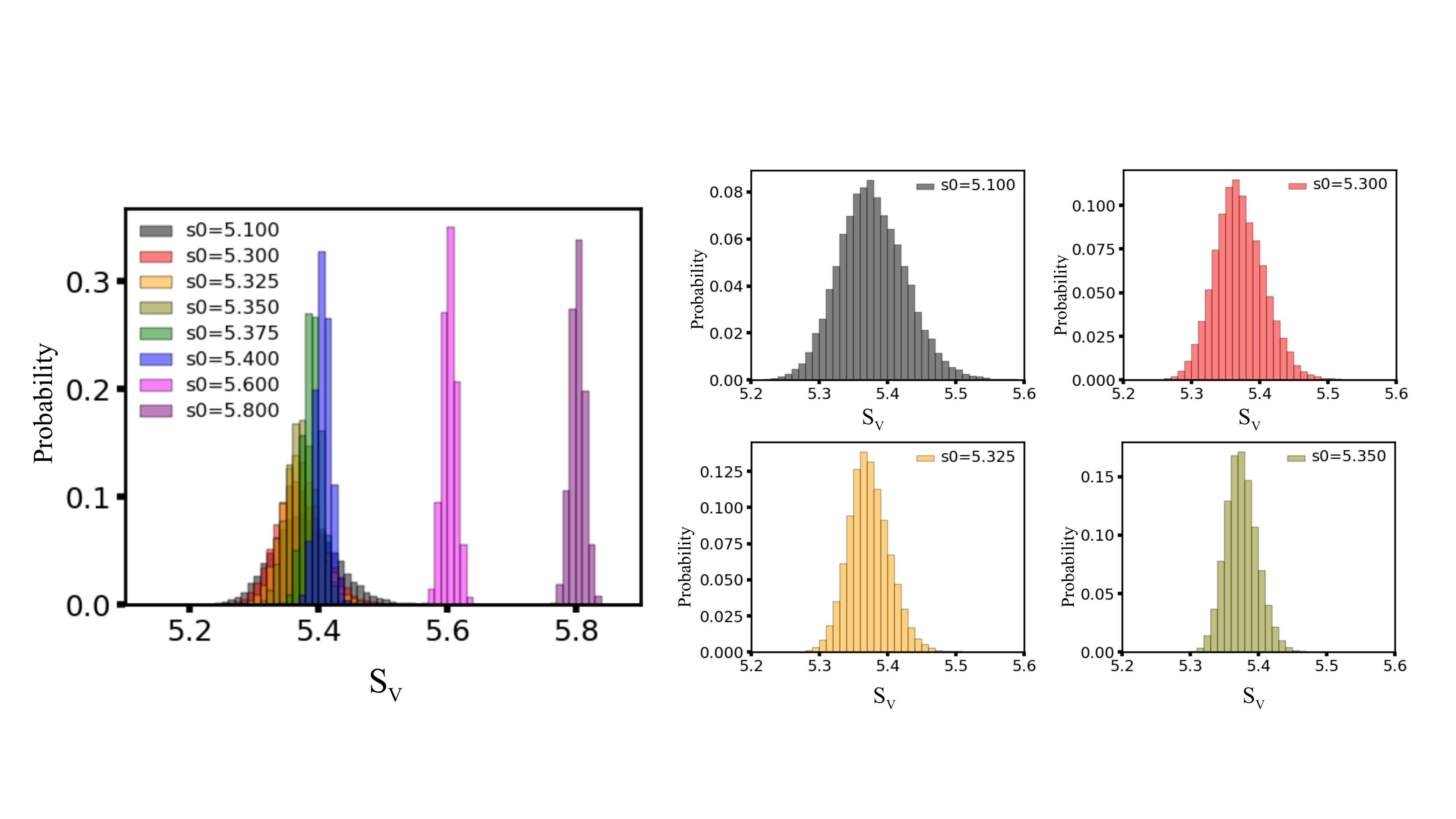}
  \caption{
  \textbf{Shape index distribution for the bulk system for different
    target shape indices.} 
 }
  \label{bulk_cell_shape}
 \end{figure*}

\newpage
\vspace{-6in}
\begin{figure*}[h]
  \includegraphics[width=14.5cm]{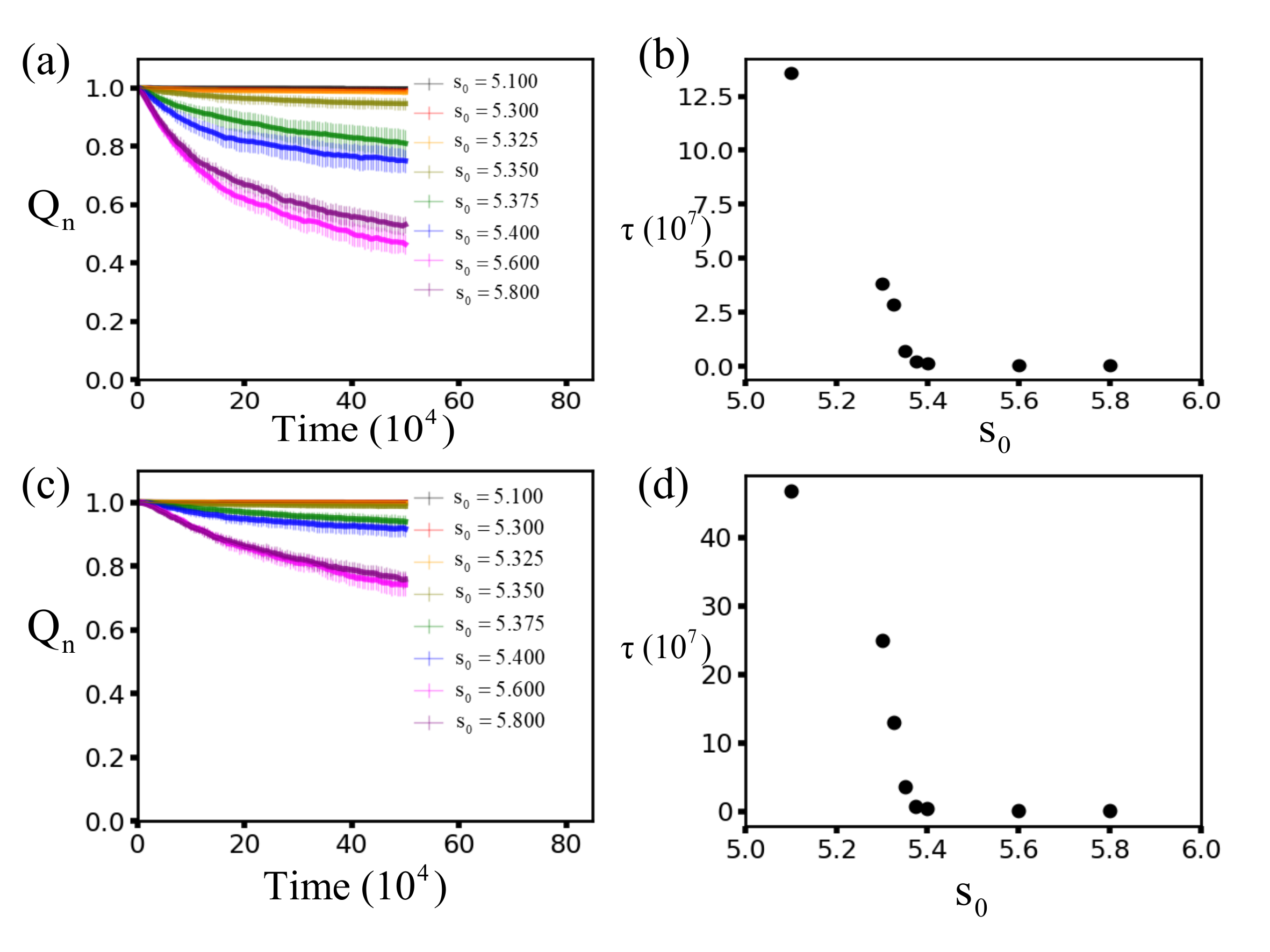}
  \caption{
   \textbf{Overlap function with extended definitions.}
   Overlap function $Q$ for different values of $s_0$ and decay time $\tau$ as a function of $s_0$, where (a, b): $w=0$ if a cell has lost three or more neighbors or (c, d): $w=0$ if a cell has lost four or more neighbors. 
 }
  \label{extension_layer}
 \end{figure*}

\newpage
\begin{figure*}[h]
  \includegraphics[width=14.5cm]{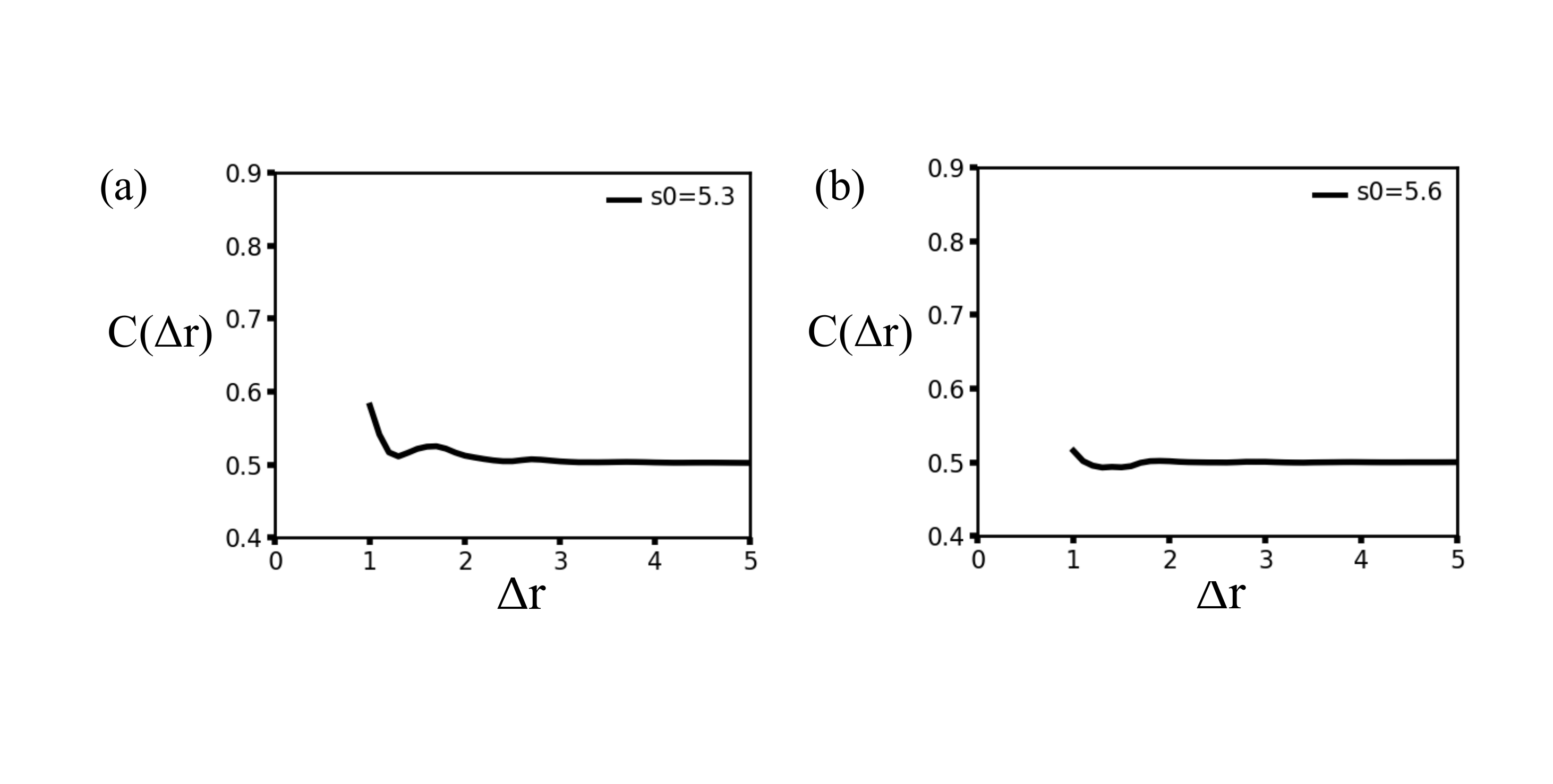}
  \caption{
  \textbf{Alignment correlation function for bulk cells.} 
  The correlation function $C(\Delta r)$ as a function of the distance $\Delta r$ between the centers of two cells 
  for the bulk system with periodic boundary conditions after equilibrium over time period $\Delta t = 110000$ 
  for (a) $s_0=5.3$ and (b) $s_0=5.6$.
 }
  \label{extension_layer}
 \end{figure*}

\newpage
\begin{figure*}[h]
  \includegraphics[width=14.5cm]{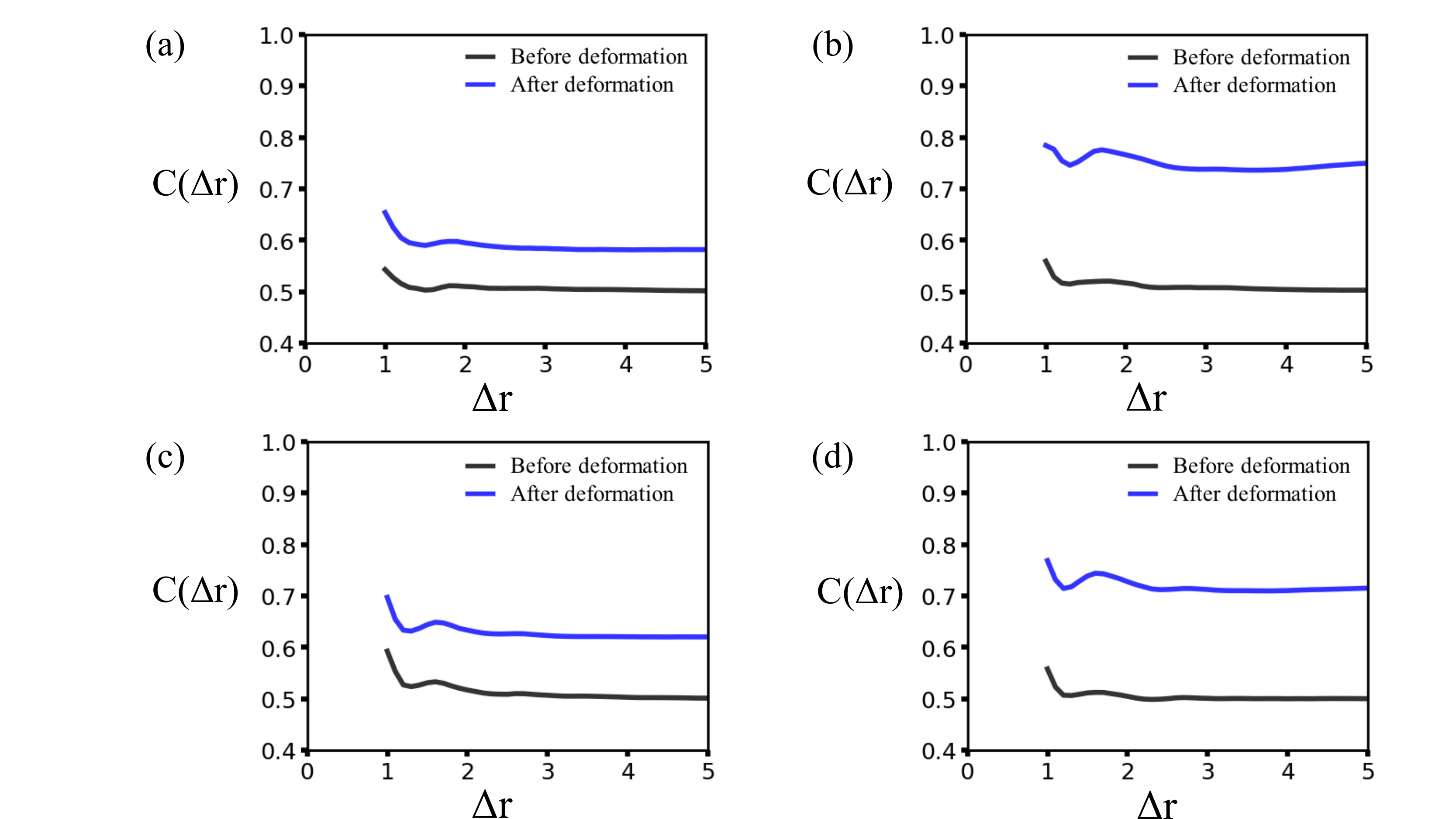}
  \caption{
  \textbf{Alignment correlation function for outermost layer and
    second outer layer.} 
  The correlation function $C(\Delta r)$ as a function of the distance $\Delta r$ between the centers of two cells 
  before and after the lateral extension 
  for (a) second outer layer cells with $s_0=5.6$; (b) outermost layer cells with $s_0=5.6$; (c) second outer layer cells with $s_0=5.0$; (d) outermost layer cells with $s_0=5.0$.
 }
  \label{extension_layer}
 \end{figure*}

\newpage
\vspace{-6in}
\begin{figure*}[h]
  \includegraphics[width=14.5cm]{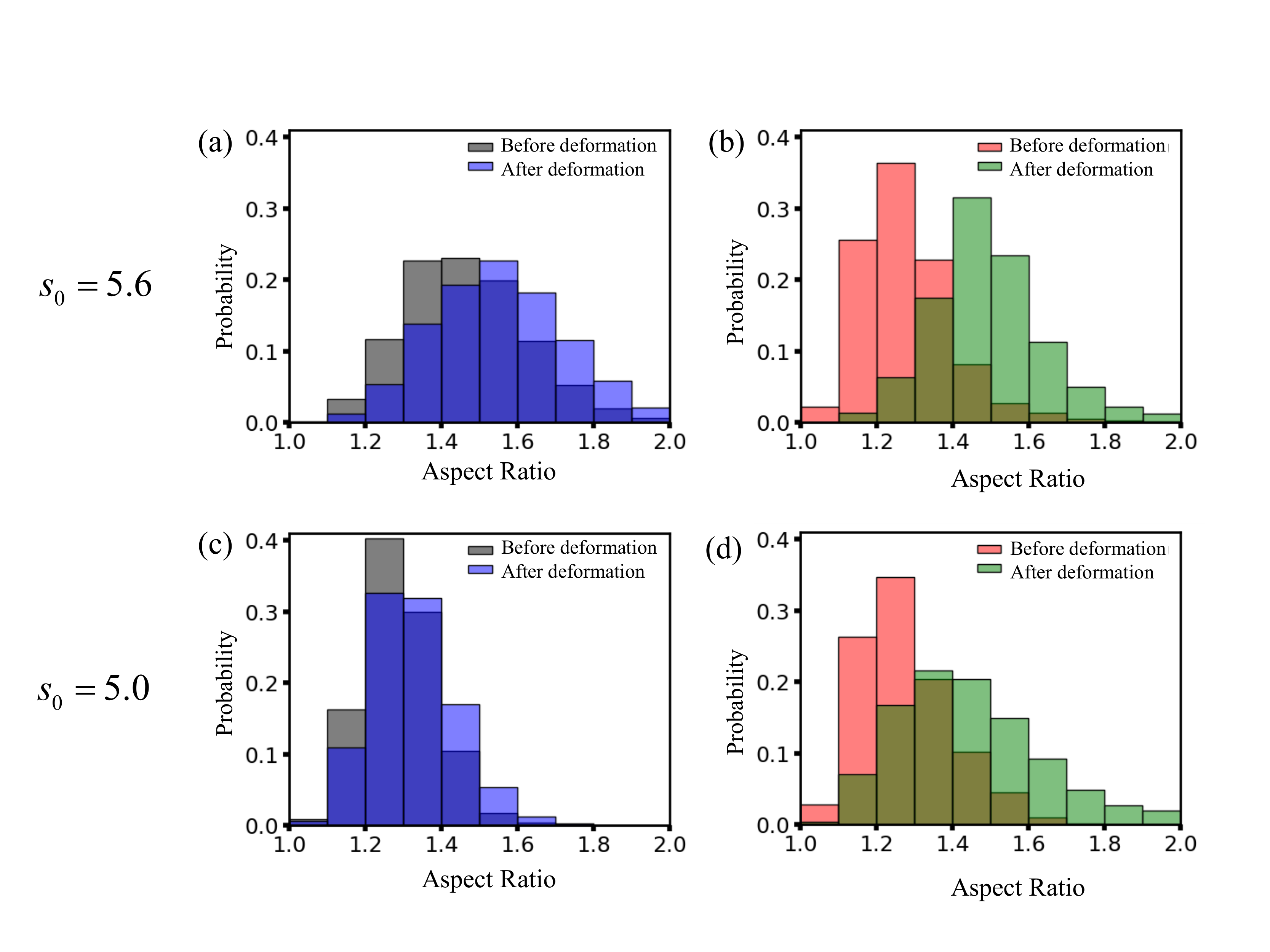}
  \caption{
  \textbf{Cell aspect ratio distributions before and after the lateral extensile deformation.} 
  We fit each cell to a minimal volume ellipsoid, and the cell aspect ratio is defined as $L_3/L_1$, where $L_1$ ($L_3$) is the shortest (longest) axis length of the ellipsoid. 
  (a) bulk cells with $s_0=5.6$; (b) boundary cells with $s_0=5.6$; (c) bulk cells with $s_0=5.0$; (d) boundary cells with $s_0=5.0$.  
 }
  \label{extension_layer}
 \end{figure*}

\newpage
\begin{figure*}[h]
  \includegraphics[width=14.5cm]{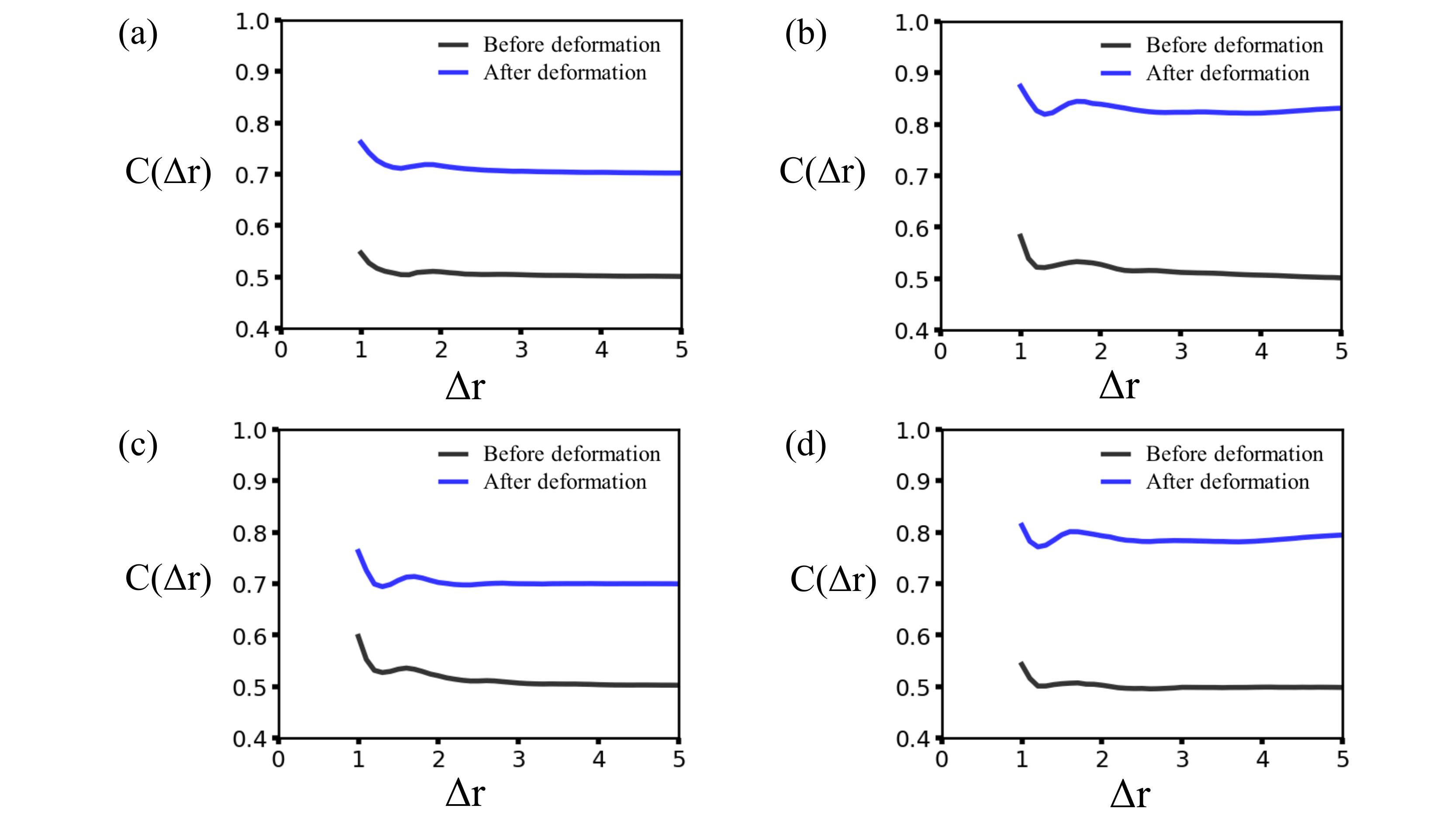}
  \caption{
  \textbf{Alignment correlation function with higher lateral extensile deformation speed.} 
  The lateral extensile deformation speed is increased to $v=10^{-3}$. The correlation function $C(\Delta r)$ as a function of the distance $\Delta r$ between the centers of two cells before and after the lateral extension 
  for (a) bulk cells with $s_0=5.6$; (b) boundary cells with $s_0=5.6$; (c) bulk cells with $s_0=5.0$; (d) boundary cells with $s_0=5.0$.
 }
  \label{extension_layer}
 \end{figure*}

\newpage
\begin{figure*}[h]
  \includegraphics[width=14.5cm]{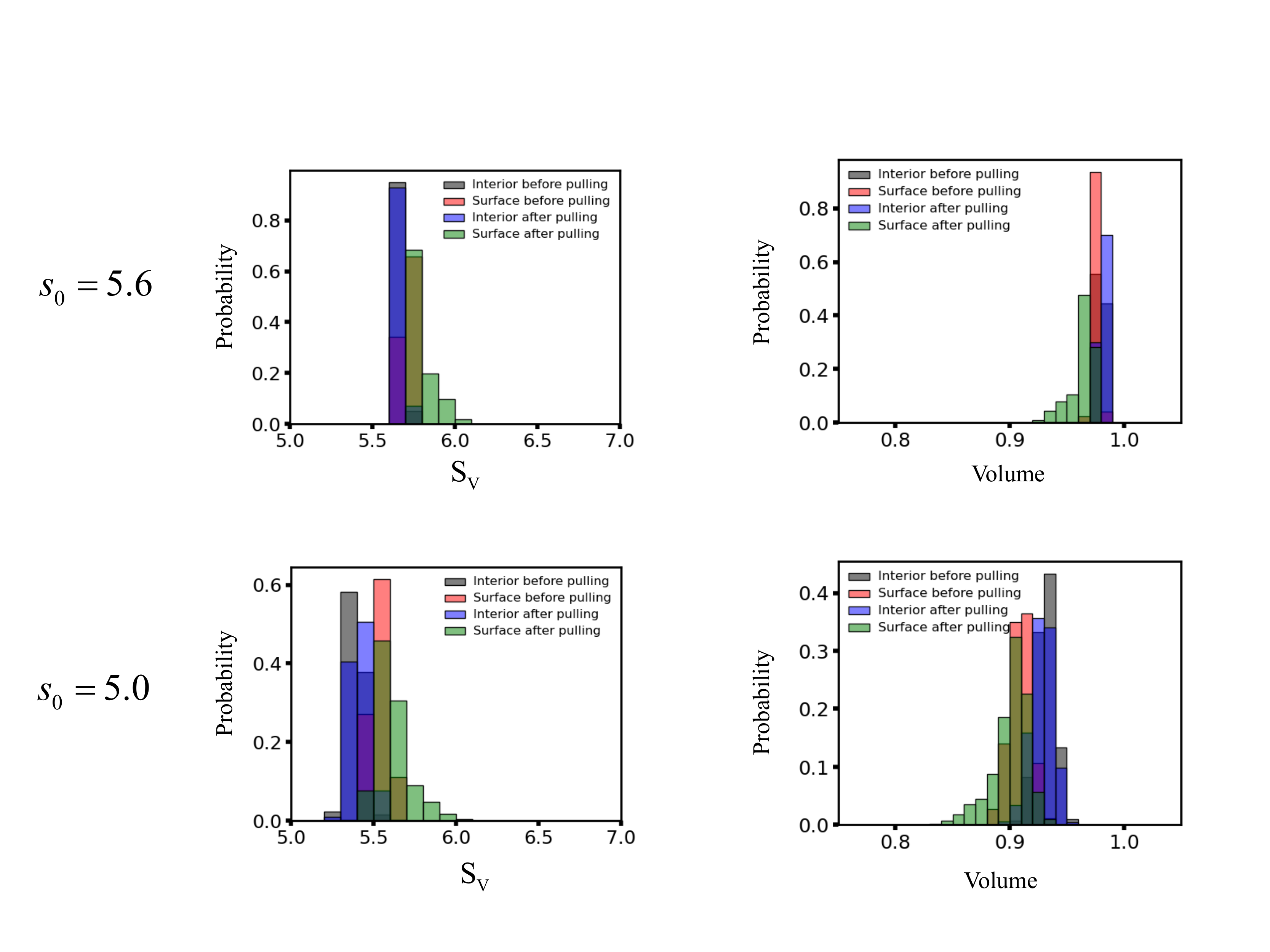}
  \caption{
  \textbf{Cell shape index and volume distributions before and after
    the lateral extensile deformation. } 
 }
  \label{extension_layer}
 \end{figure*}

\newpage
\vspace{-6in}
\begin{figure*}[h]
  \includegraphics[width=14.5cm]{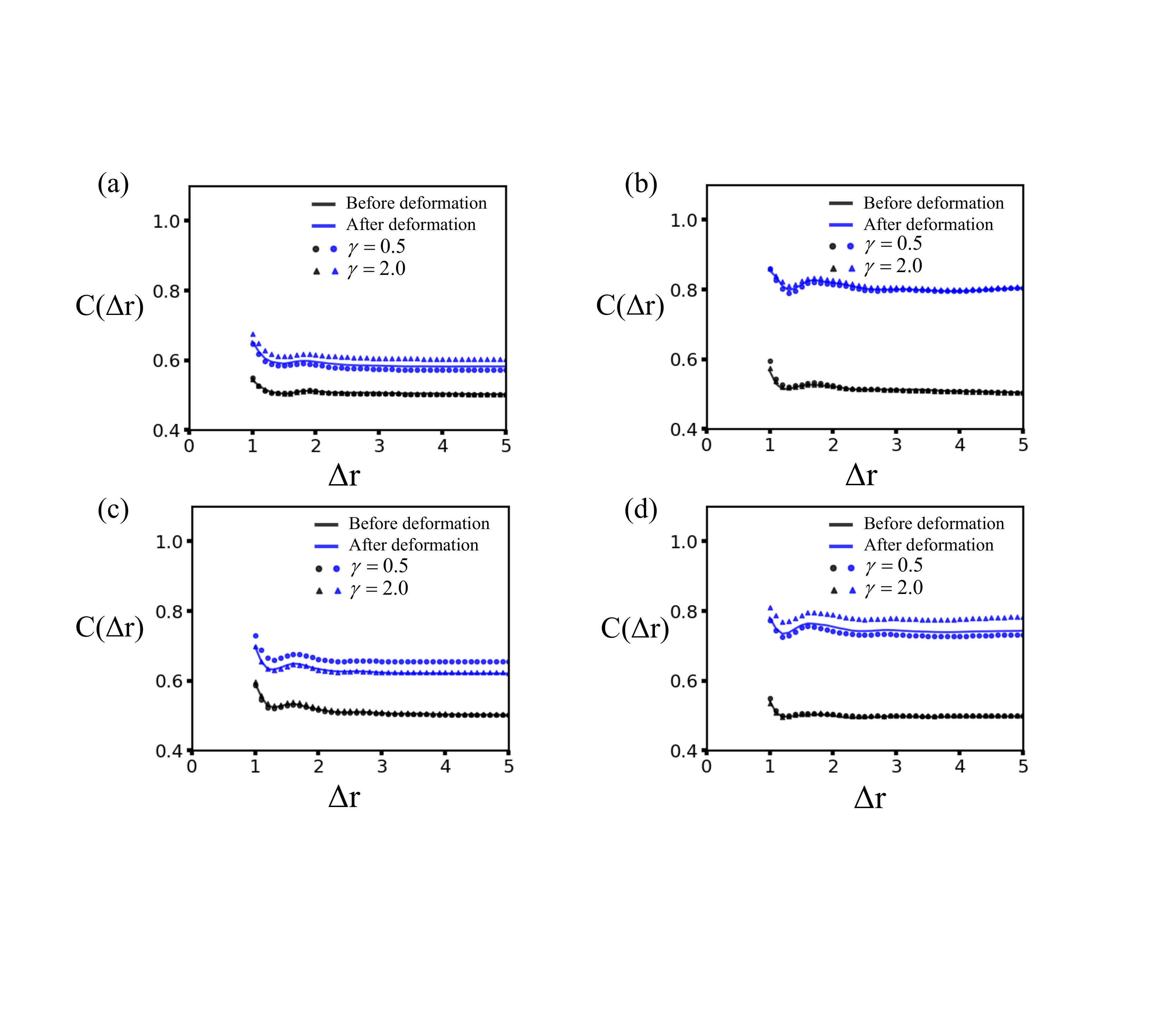}
  \caption{
  \textbf{Alignment correlation function for different values of boundary cell surface tension stiffness.} 
  The correlation function $C(\Delta r)$ as a function of the distance $\Delta r$ between the centers of two cells 
  before and after the lateral extension 
  for (a) bulk cells with $s_0=5.6$; (b) boundary cells with $s_0=5.6$; (c) bulk cells with $s_0=5.0$; (d) boundary cells with $s_0=5.0$. Curves for boundary cell surface tension stiffness $\gamma=1.0$ are shown in solid lines. Plots for $\gamma=0.5$ are shown in circular markers, and plots for $\gamma=2.0$ are shown in triangular markers. 
 }
  \label{extension_layer}
 \end{figure*}

\newpage
\vspace{-6in}
\begin{figure*}[h]
  \includegraphics[width=14.5cm]{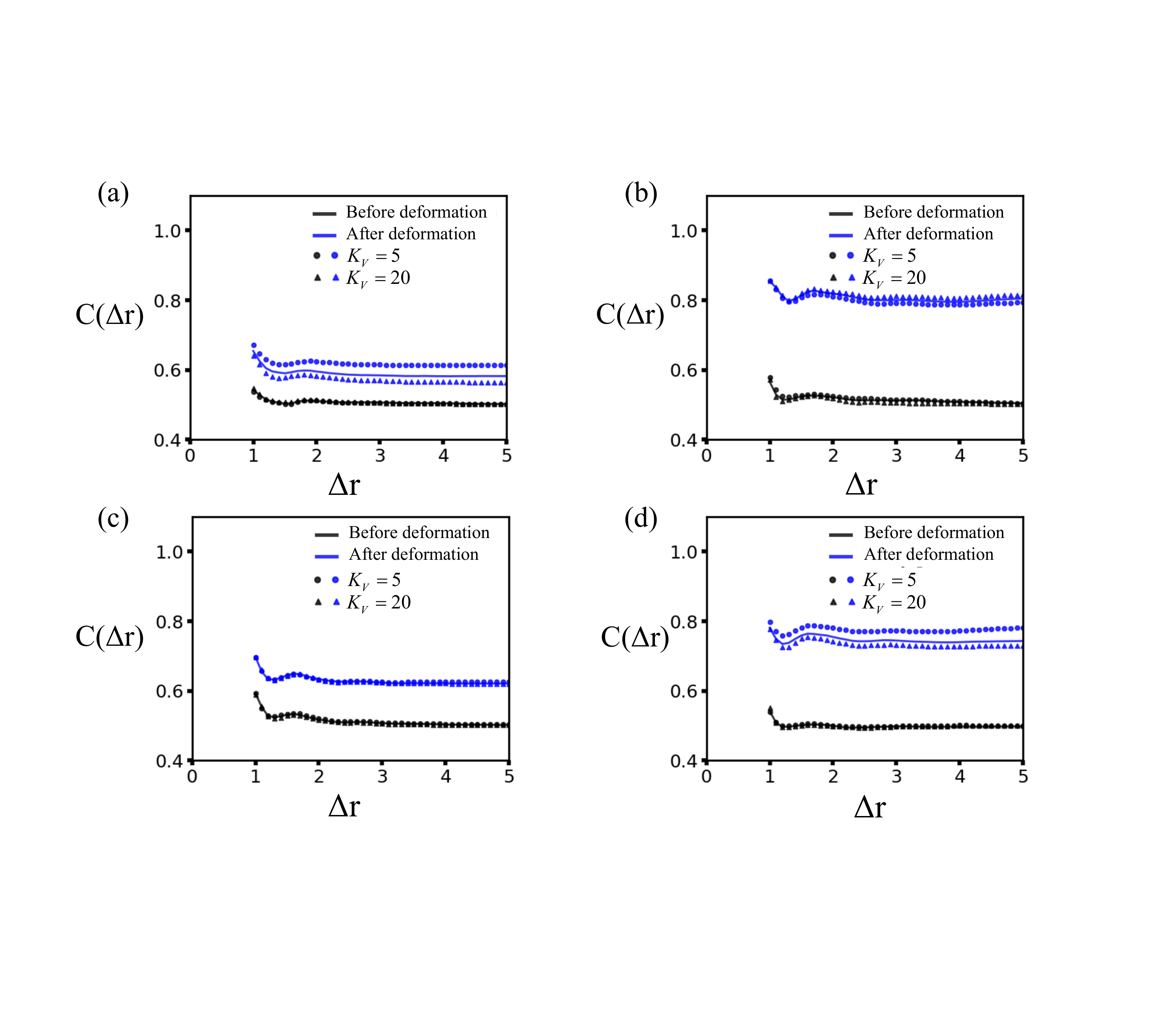}
  \caption{
  \textbf{Alignment correlation function for different values of cell volume stiffness.} 
  The correlation function $C(\Delta r)$ as a function of the distance $\Delta r$ between the centers of two cells 
  before and after the lateral extension 
  for (a) bulk cells with $s_0=5.6$; (b) boundary cells with $s_0=5.6$; (c) bulk cells with $s_0=5.0$; (d) boundary cells with $s_0=5.0$. Curves for cell volume stiffness $K_V=10$ are shown in solid lines. Plots for $K_V=5$ are shown in circular markers, and plots for $K_V=20$ are shown in triangular markers. 
 }
  \label{extension_layer}
 \end{figure*}

\newpage
\vspace{-6in}
\begin{figure*}[h]
  \includegraphics[width=14.5cm]{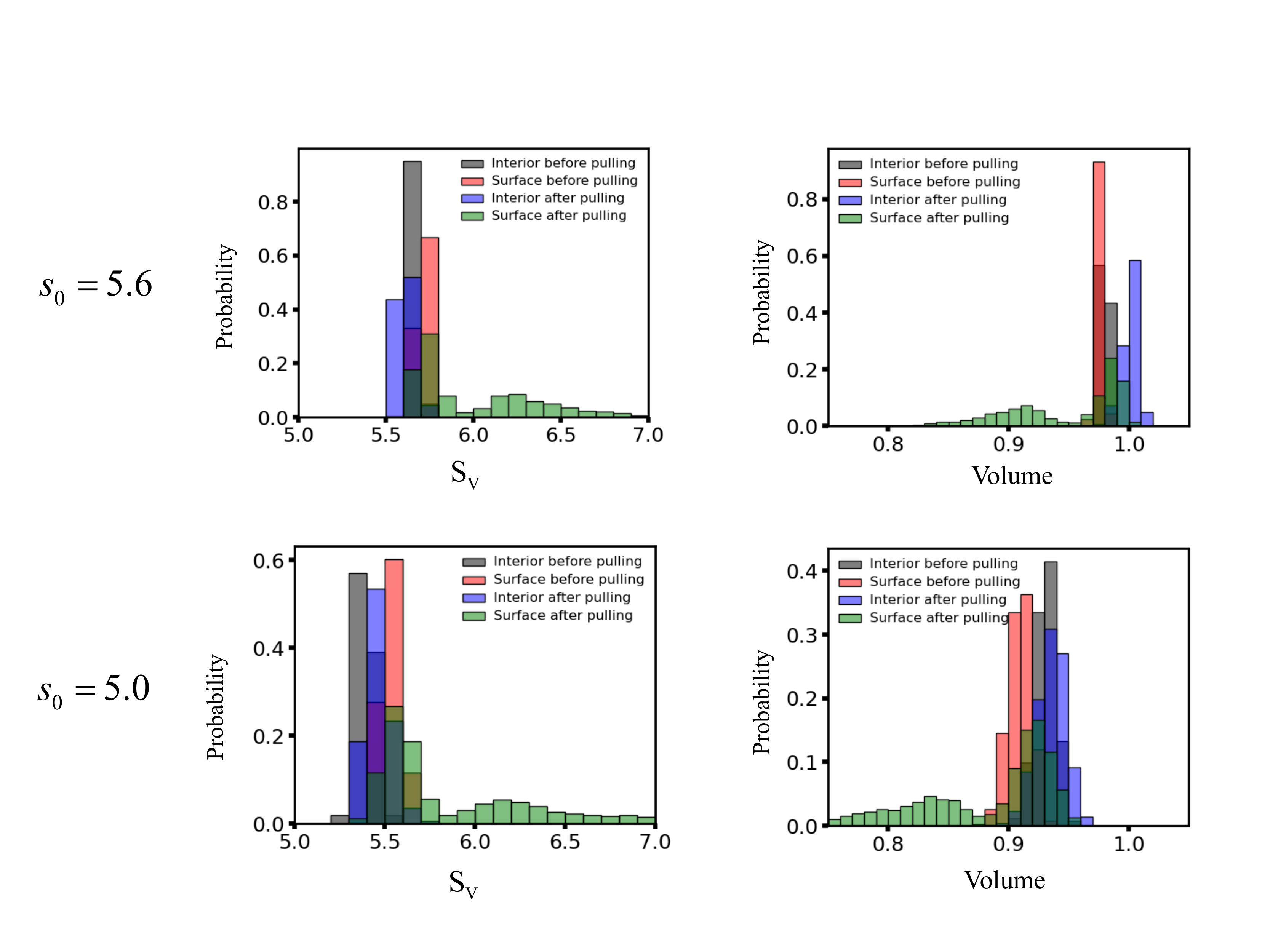}
  \caption{
  \textbf{Cell shape index and volume distributions before and after
    the radial extensile deformation. } 
 }
  \label{extension_layer}
 \end{figure*}

\end{widetext}

\clearpage

MOVIE S1. \textbf{Movie for the lateral extensile deformation of the cellular collective.} The target cell shape index $s_0=5.6$. The black colored rods indicate the orientation of cells.\\

MOVIE S2. \textbf{Movie for the radial extensile deformation of the cellular collective.} The target cell shape index $s_0=5.6$. The black colored rods indicate the orientation of cells.\\

\end{document}